\documentclass[referee]{raa}           
\usepackage{graphicx,times}
\usepackage{natbib}
\usepackage{amssymb,amsmath}
\usepackage{longtable}
\bibpunct{(}{)}{;}{a}{}{,}

\usepackage[pagebackref=true,driverfallback=dvipdfm]{hyperref}
\hypersetup{pdftitle = The title of my PDF, pdfauthor = My name, pdfsubject= The subject, pdfkeywords = keyword1 keyword2 keyword3} 
\hypersetup{colorlinks = true, linkcolor = green, anchorcolor = red, citecolor = blue, filecolor = red, urlcolor = red}

\begin{document}

\title{Six Newly Discovered Eclipsing Binary Systems in the TESS field
%$^*$
%\footnotetext{\small $*$ Corresponding author.}
}

\volnopage{ {\bf 20XX} Vol.\ {\bf X} No. {\bf XX}, 000--000}
\setcounter{page}{1}

\author{Burak Ula\c{s}$^*$\footnotetext{\small $*$ Corresponding author.} and Vildan Ayan
   %, Wen-Wen Zuo\inst{1}, Qian Yang\inst{1}, Wei-Min Yi
 %     \inst{2},  Chen-Wei Yang\inst{3,4}, Wen-Juan Liu\inst{3,4}, Peng Jiang\inst{3,4}, Xin-Wen Shu\inst{3,4}, Hong-Yan Zhou\inst{3,4}
   }

\institute{ Department of Space Sciences and Technologies, Faculty of Sciences, \c{C}anakkale Onsekiz Mart University, Terzio\v{g}lu Campus, TR~17100, \c{C}anakkale, Turkey; {\it burak.ulas@comu.edu.tr}\\
%        \and
%             Astrophysics Research Centre and Observatory, \c{C}anakkale Onsekiz Mart University, Terzio\v{g}lu Campus, TR~17100, \c{C}anakkale, Turkey\\
%	\and
%	  Center for Astrophysics, University of Science and Technology of China, Hefei 230026, China\\
%Key Laboratory for Research in Galaxies and Cosmology, The University of Science
%and Technology of China, Chinese Academy of Sciences, Hefei, Anhui, 230026, China\\
%\and 
%Polar Research Institute of China,
%Jinqiao Rd. 451, Shanghai, 200136, China\\
\vs \no
   {\small Received 20XX Month Day; accepted 20XX Month Day}
}

\abstract{We present the first detailed investigation of six eclipsing binary systems in the TESS field. The TESS light curves of the targets are analysed by determining the initial effective temperatures via SED fits. The absolute parameters are derived and the systems are compared to well-known binaries of the same type. Results show that CD-58~791, CD-62~1257 and TYC~9356-355-1 are detached binary systems. CD-54~942 and UCAC4~136-007295 are contact binaries while TYC~8508-1413-1 is a semidetached system.
\keywords{binaries: eclipsing --- stars: fundamental parameters --- stars: individual: CD-54~942, CD-58~791, CD-62~1257, TYC~8508-1413-1, TYC~9356-355-1 and UCAC4~136-007295
}
}

   \authorrunning{B. Ula\c{s} and V. Ayan}            %author_head in even pages
   \titlerunning{Six Eclipsing Binary Systems}  % title_head in odd pages
   \maketitle

%________________________________________________ sections below
% 
\section{Introduction}           %% first-level sections will be auto-capitalized
\label{sect:intro}
Studies on eclipsing binary systems improved our understanding of the structure and evolution of stars in various circumstances. Since the observation of Mizar, more than four hundred years of effort given by researchers to observe, discover, classify and reveal the characteristics of binary systems resulted in an ability to determine the light and absolute parameters, thus, making incisive estimations on their structural properties and evolutional paths. Being powerful tools for deriving the crucial properties of the stars, binary star systems still hold their solid ground as one of the leading research topics in stellar astrophysics.

Space observations are a milestone in finding a chance to look at the binary stars in a different way. They allow us to determine the variations with small amplitudes and discover very faint binaries that can be improbable to detect with numerous ground-based instruments. The discoveries made by using the data from {\it Kepler} \citep{bor10} and Transiting Exoplanet Survey Satellite (TESS, \citealt{ric15}) missions played an important role in improving our knowledge of various types of stars, as well as binary systems. They also provided a dramatic increment in the number of known systems. As of the moment, Kepler Eclipsing Binary Catalog of \cite{kir16} contains 2920 systems while \cite{pri22} cataloged 4584 binaries in their TESS Eclipsing Binary catalog.

The systems investigated in this paper were first validated as eclipsing binaries in the eclipsing binary catalog of \cite{pri22} where authors classified the targets based on their TESS light curves. The targets were also cataloged by the Gaia data releases \citep{gai16,gai18,gai22}. CD-54~942, CD-58~791 and TYC~9356-355-1 also appeared in the releases of the RAVE \citep{kor13,kun17,ste20}. \cite{sha18} listed UCAC4~136-007295 in their High Efficiency and Resolution Multi-Element Spectrograph (HERMES) catalog.

The lack of detailed information on the systems in the literature motivated us to study them in detail. Our study is the first one that focuses on the analyses of the light curves and the determination of binary parameters. The next section gives the details of the data used in the study while Sec.~\ref{sec:lcan} deals with the derivation of the initial effective temperatures, calculation of the orbital periods and the analyses of the light curves. In the last section, we conclude the results and compare the systems with the well-known binaries with the same morphologies. 

\section{Light Curves}
\label{sect:lcdat}
We collected the TESS light curve data of the systems via Mikulski Archive for Space Telescopes (MAST) portal\footnote{mast.stsci.edu/portal/Mashup/Clients/Mast/Portal.html}. The magnitude values were derived following the TESS magnitudes \citep{sta19} and using the formula $-2.5 \log F$ where $F$ is the Pre-search Data Conditioning Simple Aperture Photometry (PDC SAP) flux values stored in the data files. The data are from sectors 3, 4, 5, and 6 for CD-54~942, sectors 2,3 and 4 for CD-58~791, sector 13 for CD-62~1257, sectors 3 and 4 for TYC~8508-1413-1, sectors 1 and 13 for TYC~9356-355-1, sectors 9, 10 and 11 for UCAC4~136-007295. The TESS magnitudes are 10.785$^m$, 9.718$^m$, 9.891$^m$, 10.787$^m$, 10.238$^m$, 12.406$^m$ for CD-54~942, CD-58~791, CD-62~1257, TYC~8508-1413-1, TYC~9356-355-1 and UCAC4~136-007295, respectively. 

\begin{figure}
\centering
\includegraphics[width=\textwidth]{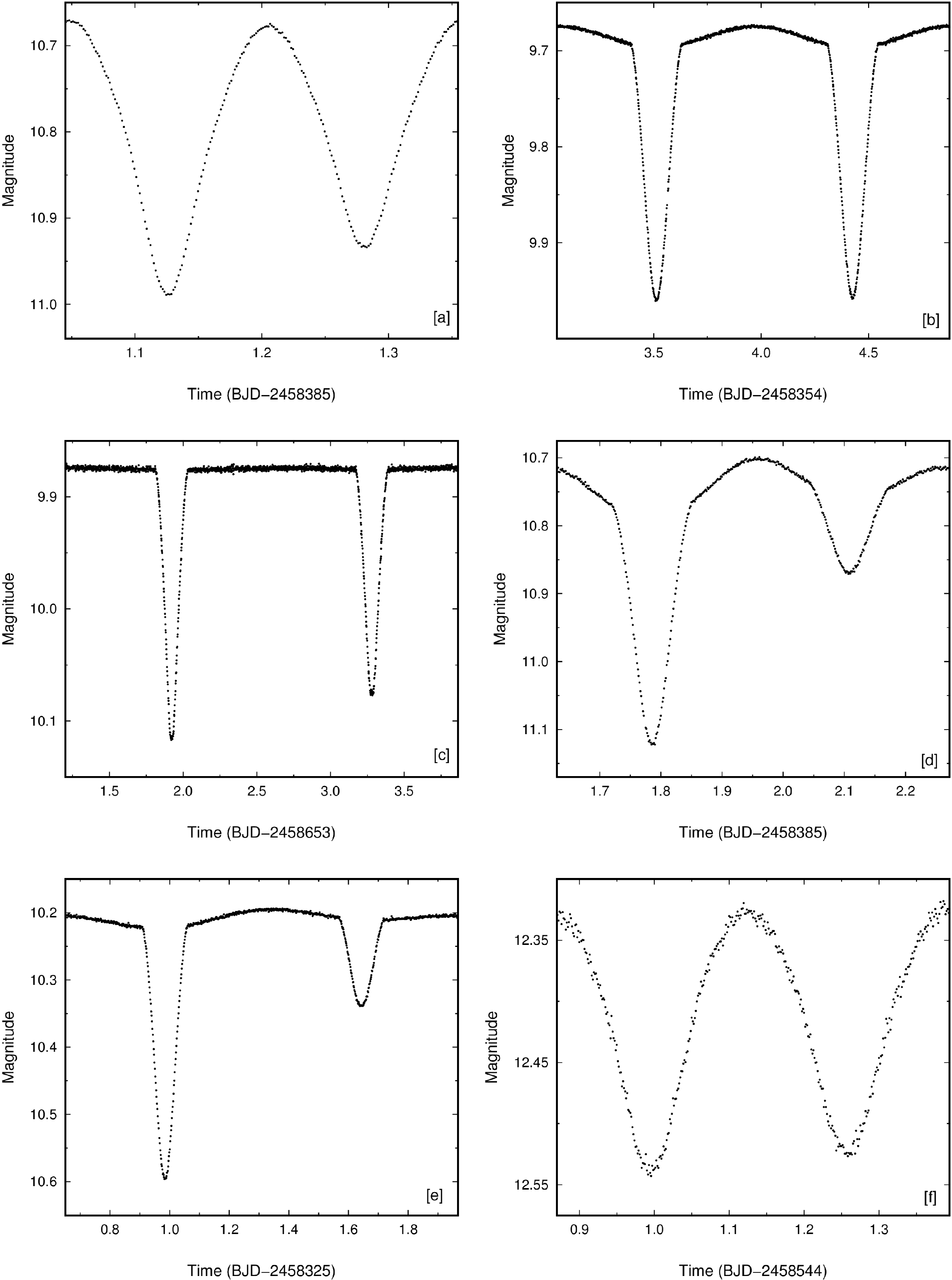}
%\includegraphics[width=0.5\textwidth]{example-image}
%\missingfigure{mag LCs}
\caption{TESS light curve in the time interval equals to one orbital period for CD-54~942 (a), CD-58~791 (b), CD-62~1257 (c), TYC~8508-1413-1 (d), TYC~9356-355-1 (e) and UCAC4~136-007295 (f).}
\label{maglc}
\end{figure}

The one orbital period long light curves of the systems are shown in Fig.~\ref{maglc}. As seen from the figure, the shapes of the curves resemble the typical variation of eclipsing binary systems in several morphologies. The depths of primary and secondary minimums are 0.31$^m$ and 0.26$^m$ for CD-54~942, 0.283$^m$ and 0.282$^m$ for CD-58~791, 0.24$^m$ and 0.20$^m$ for CD-62~1257, 0.42$^m$ and 0.16$^m$ for TYC~8508-1413-1, 0.40$^m$ and 0.14$^m$ for TYC~9356-355-1, 0.22$^m$ and 
0.21$^m$ for UCAC4~136-007295. The durations are 3.8$^h$ and 3.4$^h$, 5.5$^h$ and 5.4$^h$, 5.6$^h$ and 5.3$^h$, 2.9$^h$ and 2.8$^h$, 3.5$^h$ and 3.8$^h$, 6.1$^h$ and 5.9$^h$ for the primary and secondary minimum of CD-54~942, CD-58~791, CD-62~1257, TYC~8508-1413-1, TYC~9356-355-1 and UCAC4~136-007295, respectively. 967 times of minimum light for the targets were also calculated from the TESS light curves using the method of \citet{kwe56} and listed in Table~\ref{tabmin}

\begin{table}
\bc
\begin{minipage}[]{100mm}
\caption[]{Calculated times of minimum light for the systems. TYC~8508, TYC~9356 and UCAC4~136 refer to TYC~8508-1413-1, TYC~9356-355-1 and UCAC4~136-007295, respectively. This table is for guidance, its full version is available in electronic form.}\label{tabmin}\end{minipage}
\setlength{\tabcolsep}{1pt}
\small
 \begin{tabular}{lccc}
  \hline\noalign{\smallskip} 
 System~ & ~BJD-2457000 & Error & Min. type \\
\hline   
CD-54~942 &~~1386.126517&~~0.000001&I\\
CD-54~942 &~~1385.971296&~~0.000001&II\\
CD-54~942 &~~1386.436137&~~0.000001&I\\
CD-54~942 &~~1386.281039&~~0.000001&II\\
CD-54~942 &~~1386.745805&~~0.000001&I\\
CD-54~942 &~~1386.590640&~~0.000001&II\\
\multicolumn{1}{c}{$\vdots$}& $\vdots$ & $\vdots$ &$\vdots$\\
\hline
\end{tabular}
\ec
\end{table}

\section{Analyses of the Light Curves}
\label{sec:lcan}
\begin{figure}
\centering
\includegraphics[width=\textwidth]{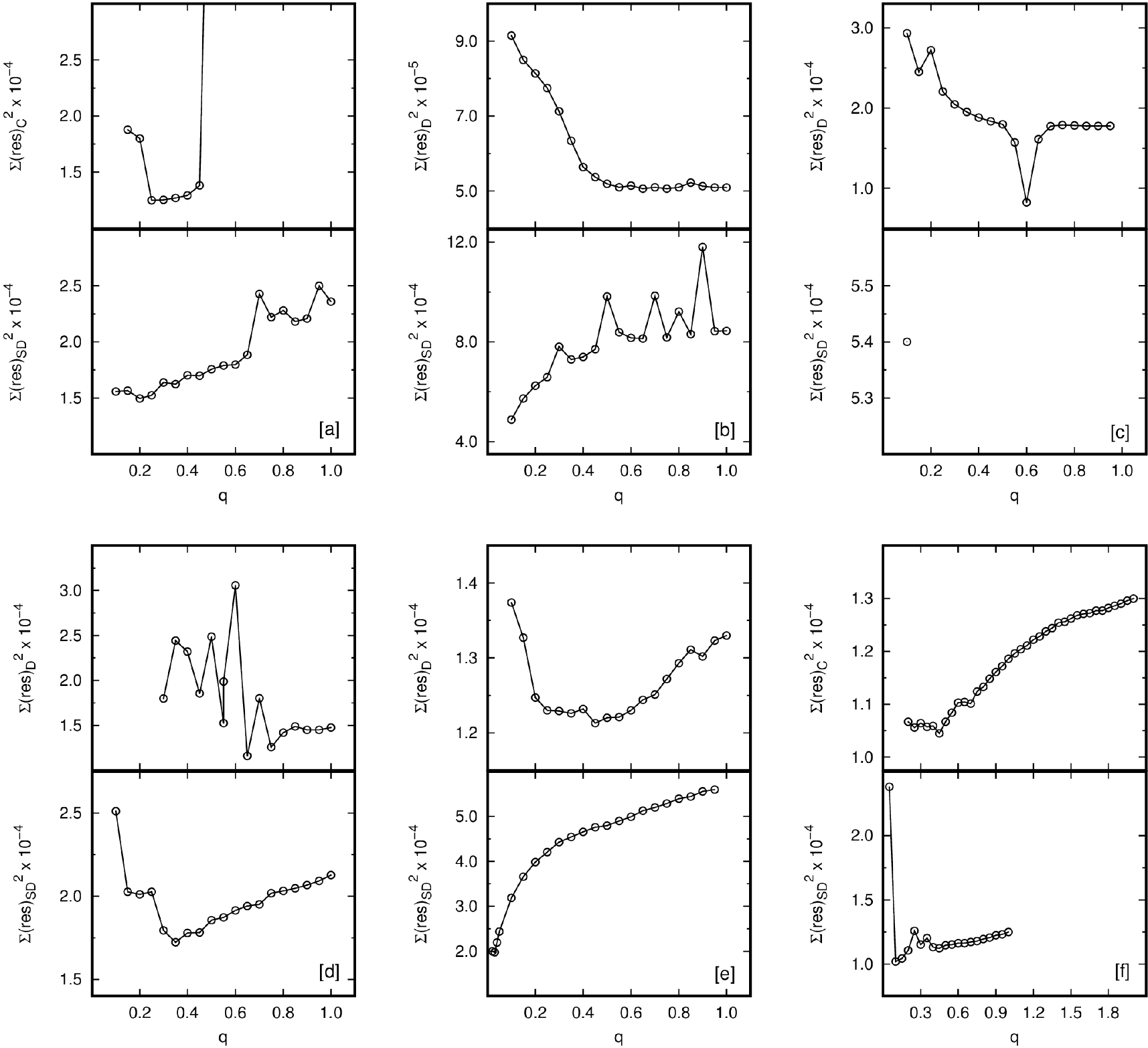}
%\includegraphics[width=0.5\textwidth]{example-image}
%\missingfigure{q search}
\caption{Results of the $q$-search for CD-54~942 (a), CD-58~791 (b), CD-62~1257 (c), TYC~8508-1413-1 (d), TYC~9356-355-1 (e) and UCAC4~136-007295 (f). The sub-indices C, D and SD indicate contact, detached and semidetached morphologies, respectively. Note that the procedure output unphysical results for CD-62~1257 in semidetached assumption other than the initial value.}
\label{qsearch}
\end{figure}

The systems were recently classified as eclipsing binary candidates by \cite{pri22}, however, their binary parameters were not determined in the literature. The lack of detailed information about the systems necessitates the derivation of some crucial parameters prior to the light curve analyses. We first determined the mass ratio for the systems via $q$-search procedure by using the 2015 version of Wilson-Devinney code \citep{wil71,wil20} on 1000 normalized data points of the light curves. A lack of morphological classes of the systems directed us to apply $q$-search in two different morphological assumptions for each system. Contact and semidetached approach used for CD-54~942 and UCAC4~136-007295. The q-searches with the contact approach were applied in the range of $q$=0.0-2.0 to avoid missing the more accurate value larger than 1.0. The process output unphysical results for the q values larger than 0.5 for CD-54~942. The q-search for CD-58~791, CD-62~1257, TYC~8508-1413-1 and TYC~9356-355-1, on the other hand, was made in detached and semidetached mode. The results of the $q$-search are shown in Fig.~\ref{qsearch} which indicates that CD-54~942 and UCAC4~136-007295 are probably contact binary systems whereas detached morphology is more accurate for CD-58~791, CD-62~1257 and TYC~9356-355-1. TYC~8508-1413-1 are found to be semidetached binary based on our $q$-search results. However, one must bear in mind that the light curve is not sensitive to mass ratio for detached binaries and semidetached systems that show partial eclipses \citep{ter05}, therefore, some further properties must be taken into account to decide the morphology as we remark in corresponding subsections.

The effective temperature also significantly impacts deriving physically meaningful results by light curve analysis. The systems were not studied in detail previously, thus, we faced the lack of reliable effective temperature values for the components in the literature as expected. We constructed spectral energy distributions (SEDs) through Virtual Observatory SED Analyzer \citep[VOSA,][]{bay08} using the photometric data given by the VizieR database \citep{och00} to derive the temperatures for the components of the systems. During the SED analysis, the data were fitted using chi-squared fit or binary fit (two components approach) with a parameter-grid search to obtain the optimized Kurucz atmosphere model \citep{kur79}. The search domain for the temperatures was set between 3500 and 10000~K while the interval for log~$g$ values, which has a minor effect on SED, was between 2.5-5.0. A good fit corresponds to the Vgf$_b$ value, a value for estimating the goodness of fit, that is smaller than 10-15 \citep{bay08}. The error in the temperature and the gravity values calculated by the SED analyses are $\Delta T$=125~K and $\Delta$ log~$g$=0.25. The results of the SED analyses are given in corresponding subsections while ~the fits are shown in Fig.~\ref{sed}.

Orbital periods of the systems were determined by calculating averages of the differences between two consecutive minima times of the same type. Frequency analyses using {\tt Period04} software \citep{len05} were also applied to the light curves to confirm the calculated orbital period values. The calculated times of minima (Table~\ref{tabmin}) were also used for a period analysis with the linear variation assumption and variation in orbital period $\Delta P$ and time of minimum light, $\Delta T_0$ were derived utilizing the equation $O - C = \Delta T_{0} + \Delta P  \cdot E$,
with the cycle number $E$. The standard errors ($SE$) for the orbital periods were derived by using the equation containing standard deviation ($\sigma$) and the number of derived period values ($N$), $SE = \sigma / N$. The calculated values are given in the following subsections as well as Table~\ref{tablc}.

The analyses were applied by using PHOEBE \citep{pri05} which analyse the input light curve by using Wilson-Devinney method \citep{wil71}. The inclination $i$, the temperature of the secondary component $T_2$, mass ratio $q$, the surface potential values of the primary component $\Omega_1$ and the luminosity of the primary component $L_1$  were the adjustable parameters of the solutions. $\Omega_2$ was also set as adjustable in the analyses with detached approximation. The albedos ($A_{1}$, $A_{2}$) were adopted from \cite{ruc69} and the gravity darkening coefficients ($g_{1}$, $g_{2}$) for the systems were taken from \cite{zei24} and \cite{luc67} by taking account of the fact that the granulation boundary at about F0 spectral type (\citealt{gra89}). Logarithmic limb darkening coefficients ($x_{1}$ and $x_{2}$) were derived from \cite{cla17} based on the initial temperatures of the components. The light curve data covering more than 50000 data points were reduced to 50000, the input limit of the software, during the solution.

\subsection{CD-54~942}

\begin{figure}
\centering
\includegraphics[width=\textwidth]{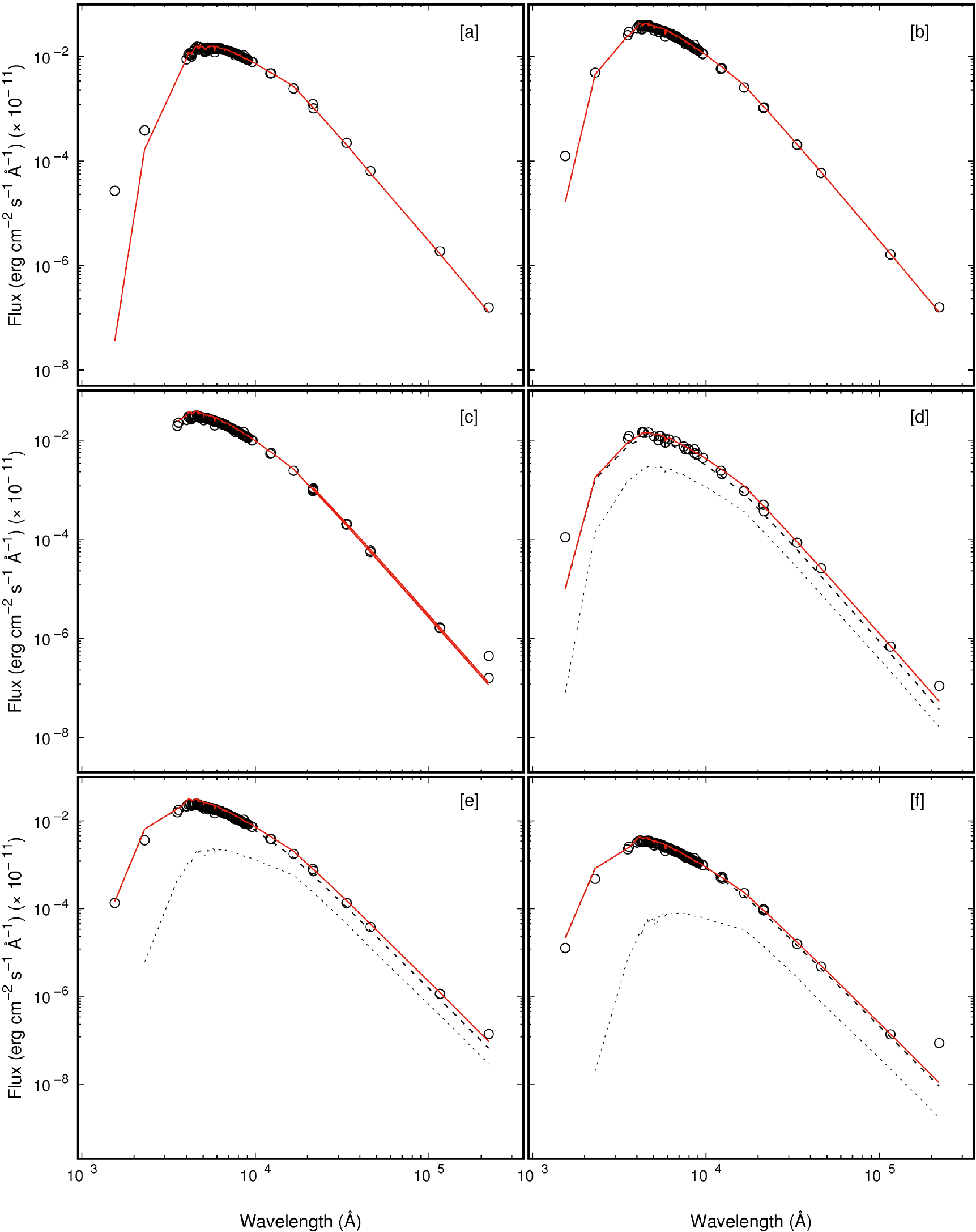}
%\includegraphics[width=0.5\textwidth]{example-image}
%\missingfigure{q search}
\caption{SED fits applied to the available photometric data for CD-54~942 (a), CD-58~791 (b), CD-62~1257 (c), TYC~8508-1413-1 (d), TYC~9356-355-1 (e) and UCAC4~136-007295 (f). The red lines refer to the model fits. The dashed and dotted lines in d, e and f represent the best model fits for the hotter and the cooler components, respectively.}
\label{sed}
\end{figure}

The analysis was applied to 50000 data points which are binned from originally 59124. The orbital period was set to 0.309691 days, the value derived by using the previously discussed method. Binary fit in the SED analysis resulted in the effective temperatures of the two components as 7250 and 5000~K. However, the first one is quite large compared to the values 5502~K \citep{ste20}, 5244~K \citep{gai16,gai22} and 5241 \citep{sta19} given in the literature. Therefore, we applied a chi-square fit to determine a more realistic temperature and derived the value of 5250~K. The log~$g$ was also found to be 4.5. The distance is 157.4(3) based on the Gaia DR3 parallax \citep{gai16,gai22} and the extinction is 0$\overset{m}{.}$0342, derived by using the Galactic Dust Reddening and Extinction interface of NASA/IPAC Infrared Science Archive\footnote{\url{https://irsa.ipac.caltech.edu/applications/DUST/}}. The components are assumed to be in solar abundance. The V$_{gfb}$ of the resulting fit is 0.704, which is agree with the criterion for the good fit. The result is shown in Fig.\ref{sed}.

\begin{figure}
\centering
\includegraphics[width=\textwidth]{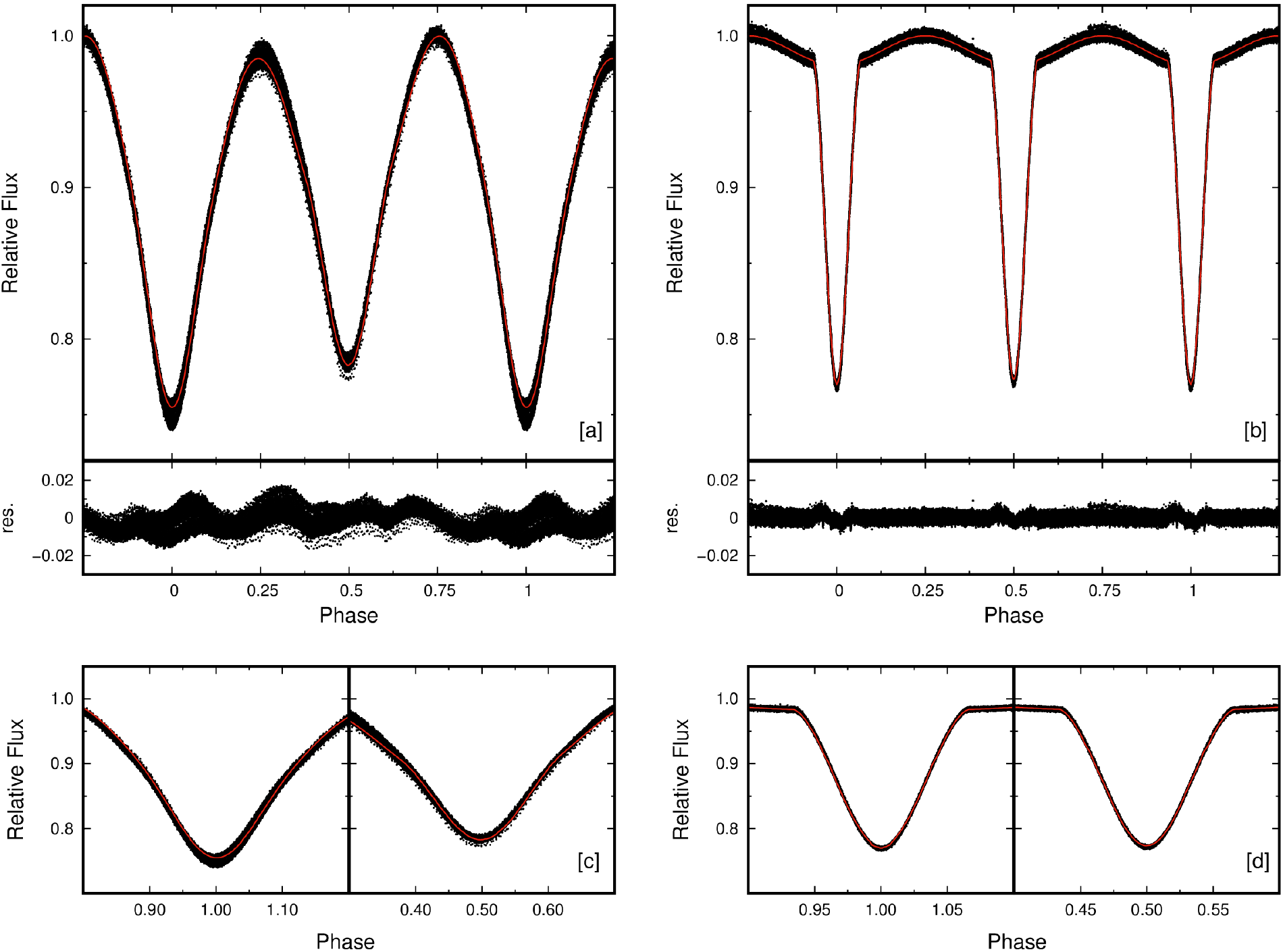}
\caption{Calculated light curve (red lines) is plotted with the observations and the agreement at the minimum phases are represented for CD-54~942 (a, c) and CD-58~791 (b, d).}
\label{lcs1}
\end{figure}

The effective temperature of the hotter component is adjusted to 5250~K during the light curve solution following the SED analysis. The light curve solution was conducted in contact and semidetached approximations with the initial value of the mass ratios of 0.25 and 0.2 considering the $q$-search. A significantly better fit with smaller squared residuals was achieved with the contact approach. Additionally, a spotted area on the cooler components was hypothesized to fit the difference between the maxima of the light curve. The parameters of the spot are found to be $\beta$=45$^{\circ}$, $\lambda$=90$^\circ$, $r$=25$^\circ$ and $t$=0.9, where $\beta$, $\lambda$, $r$ and $t$ are co-latitude, longitude, fractional radius and temperature factor, respectively. The results of the analysis are listed in Table~\ref{tablc} while the agreement of the final fit to the observations is shown in Fig.~\ref{lcs1}. We conclude that the target can be considered as a contact binary system.

\begin{table}
\bc
\begin{minipage}[]{100mm}
\caption[]{Results of the light curve analyses. TYC~8508, TYC~9356 and UCAC4~136 refer to TYC~8508-1413-1, TYC~9356-355-1 and UCAC4~136-007295, respectively. The standard deviations, 3$\sigma$ for the last digits of light parameters are given in parentheses. The errors in the effective temperature, $T_1$, are from the corresponding SED analyses. See text for the error determination for orbital period values, $P$.}\label{tablc}\end{minipage}
\setlength{\tabcolsep}{1pt}
\small
 \begin{tabular}{lccc}
  \hline\noalign{\smallskip}
Parameter 		& CD-54~942 		& CD-58~791 		& CD-62~1257 \\
\hline           
T$_0$ (BJD) 		& 1435.365982(3)~ 	& 1355.68657(2)~ 	& 1654.92067(2)~ 	\\    
$P$~(days)   		& 0.309691(4)  		& 1.82630(2) 		& 2.71459(4)  	\\                                       
$i$ ${({^\circ})}$      & 66.81(2)   		& 77.922(4) 		& 83.41(1) 		\\       
$q$                     & 0.237(1) 		& 0.789(1) 		& 0.608(1)		\\         
$T_1$ (K)               & 5250(125) 		& 6500(125) 		& 6500(125) 		\\   
$T_2$ (K)               & 4963(21)		& 6448(10)		& 6273(33)		\\           
$\Omega _{1}$           & 2.293(1) 		& 5.022(2)		& 6.630(6) 		\\         
$\Omega _{2}$           &$\Omega _{1}$ 		& 4.896(2)		& 7.90(1)		\\          
${L_1}/({L_1 +L_2})$  & 0.832(1) 		& 0.5701(4)		& 0.804(1)		\\  
$r_1$                   & 0.5164(6) 		& 0.2381(7)		& 0.166(2) 		\\    
$r_2$                   & 0.271(3) 		& 0.2097(9)		& 0.091(4)		\\ 
$x_{1},~x_{2}$ 		& 0.646, 0.659 		& 0.558, 0.558 		& 0.568, 0.581		\\
\hline
			& TYC~8508 		& TYC~9356		& UCAC4~136 \\
\cline{2-4}
T$_0$ (BJD) 		& 1386.143736(7)~ 	& 1325.984991(7)~ 	& 1544.00685(2)~ 	\\    
$P$~(days)   		& 0.64191(3)  		& 1.31699(3) 		& 0.52391(8)  	\\                                       
$i$ ${({^\circ})}$      & 69.86(2)   		& 79.67(3) 		& 67.46(3) 		\\       
$q$                     & 0.580(1) 		& 0.517(1) 		& 0.102(1)		\\         
$T_1$ (K)               & 6500(125) 		& 7000(125) 		& 7000(125) 		\\   
$T_2$ (K)               & 5058(36)		& 5548(21)		& 6682(51)		\\           
$\Omega _{1}$           & 3.848(6) 		& 5.244(3)		& 1.956(1) 		\\         
$\Omega _{2}$           & $\Omega _{cr}$=3.027	& 4.137(3)		&$\Omega _{1}$		\\          
${L_1}/({L_1 +L_2})$  & 0.743(1) 		& 0.796(1)		& 0.905(2)		\\  
$r_1$                   & 0.311(3) 		& 0.212(1)		& 0.583(1) 		\\    
$r_2$                   & 0.331(1) 		& 0.179(2)		& 0.210(6)		\\ 
$x_{1},~x_{2}$ 		& 0.570, 0.617 		& 0.547, 0.670 		& 0.548, 0.601		\\
\hline
\end{tabular}
 \ec

\end{table}

\subsection{CD-58~791}
The SED analysis for the system resulted in two temperature values, 7750 and 6500~K, using the binary fit approach. However, the former temperature are contrary to the values given by \cite{gai16,gai22}, \cite{sta19} and \cite{kor13} which are between 6000 and 6600~~K. Therefore, we decided to analyse the SED by using the chi-square fit option. The result indicated that log~$g$=4.0 with the temperature value of 6500~K which we set as the temperature of the primary component in our light curve analysis. The extinction was set to 0$\overset{m}{.}$0736 and the distance was taken as 416(2)~pc following \citep{gai16,gai22}. The V$_{gfb}$ of the solution, 0.81, corresponds to a good fit. The SED fit is plotted in Fig.~\ref{sed}. The orbital period of the system, on the other hand, was calculated as 1.82630 by following the method described earlier.

The light curve analyses were conducted on 44540 data points in two morphologies, detached and semidetached, individually. Although the semidetached configuration shows relatively small squared residuals in the min of $q$-search curve, the detached configuration fits the observations better in the light curve analysis. The reason may rise from the fact that the mass ratio is not sensitive to the light curve in detached configuration and weakly depend on it in the semidetached binaries showing partial eclipses \citep{ter05}. The results are shown in Table~\ref{tablc} as well as in Fig.~\ref{lcs1}. Considering the above-mentioned arguments, we accomplish that the system is a detached binary.

\subsection{CD-62~1257}
The temperature of the primary components was derived in the same method that was described in previous subsections. The chi-square fit resulted in log~$g$=4.5 and T=6500~K which is agree with the values given by \cite{gai16,gai22} and \cite{sta19}. The distance of the system was set to 455(7) pc based on the parallax by\cite{gai16,gai22}, thus, the extinction is 0$\overset{m}{.}$1629. The V$_{gfb}$ was derived as 0.577, which quite agrees with the acceptable interval. Furthermore, the orbital period of the system was calculated to be 2.71459~days using the method discussed above.

\begin{figure}
\centering
\includegraphics[width=\textwidth]{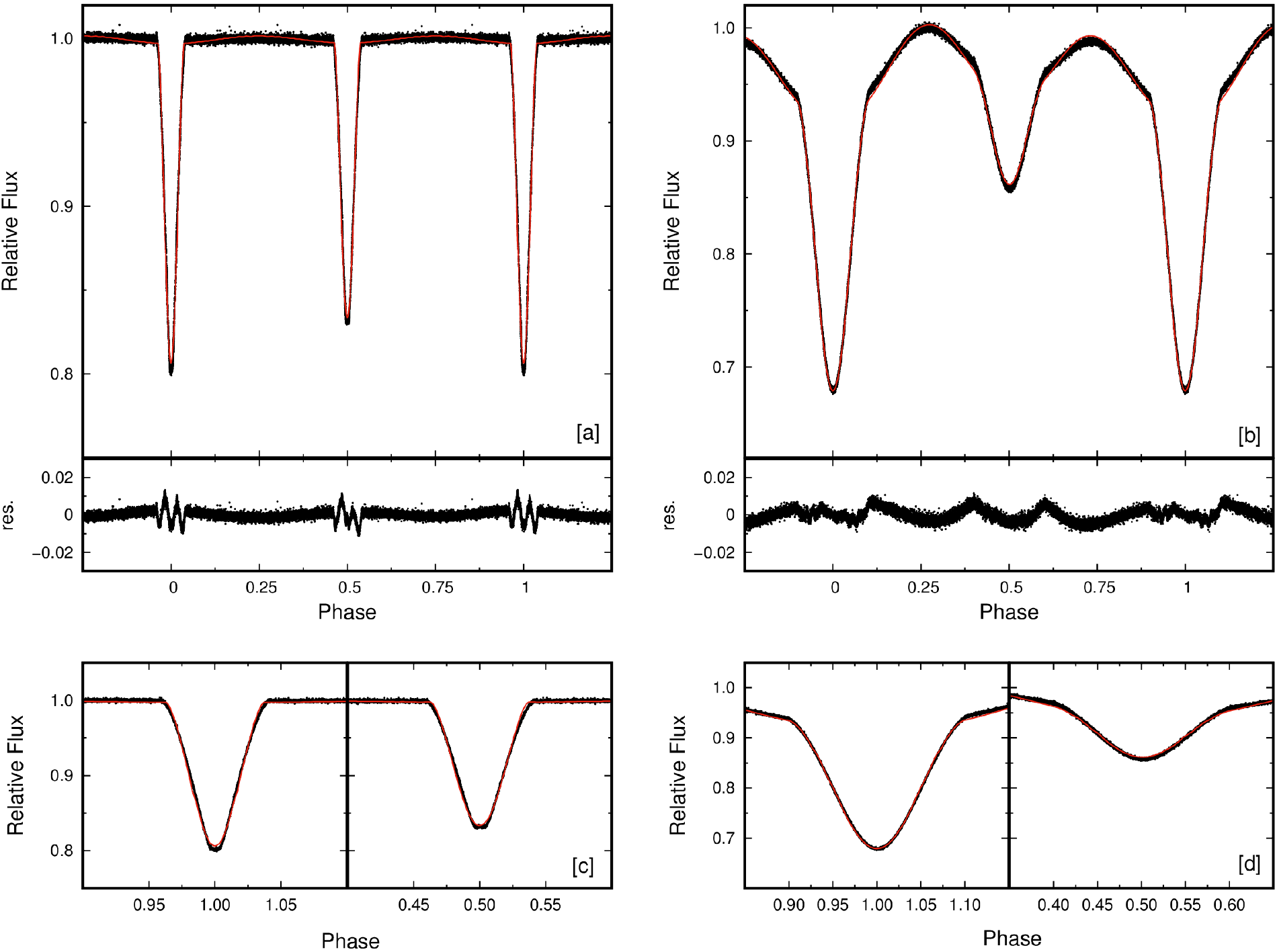}
%\includegraphics[width=\textwidth]{fig_pdf/lc_3871_1980_b.pdf}
%\missingfigure[figwidth=\textwidth]{lc 3871 1980}
\caption{Same as \ref{lcs1}, but for CD-62~1257 (a, c) and TYC~8508-1413-1 (b, d).}
\label{lcs2}
\end{figure}

The light curve of the binary was analysed by assuming that the system is a detached binary considering not only the $q$-search but also the period value and the shape of the light curve. The number of data in the solution was 19581. The calculated light curve is shown with the observations in Fig.~\ref{lcs2} while the results are listed in Table~\ref{tablc}.

\subsection{TYC~8508-1413-1}
SED analysis applied to available photometric data by using binary fit which resulted in $T_1$=6500~K, $T_2$=5750~K and log~$g_1$=log~$g_2$=4.5, the values assumed as the initial parameters for the components during the light curve analysis. The V$_{gfb}$ of the fit was calculated as 4.81, smaller than 10-15. The distance of the system was adopted as 374(2)~pc \citep{gai16,gai22} that corresponds to the extinction value of $A_V$=0$\overset{m}{.}$04. 

The analysis was applied to 12399 data points with detached and conventional semidetached approaches individually. Although the residuals of the light curve solution were close in two morphologies, the fillout factor for the secondary component was calculated to be 99\%, using the equation, $1- [(\Omega_{2} - \Omega_{cr}) / \Omega_{cr}]$ in detached approximation, where $\Omega_{cr}$ denotes the critical potential value for the corresponding mass ratio. In addition, the K spectral type of the secondary is typical for the secondaries of the Algol-type binaries. Therefore, the semidetached configuration was supposed to be a more reliable morphology of the system. The initial mass ratio was set to 0.35 and the orbital period was taken as 0.64191~days during the solution. The albedo of the secondary component was set as a free parameter for the first couple of runs in order to achieve a better fit around the secondary minimum. A hot spotted area on the primary component, which may have been the result of the mass transfer from the secondary, was also included in the solution to fit the difference between levels at maximum phases. The spot parameters are $\beta$=90$^{\circ}$, $\lambda$=240$^\circ$, $r$=15$^\circ$ and $t$=1.05.  The resulting parameters are listed in Table~\ref{tablc}. The calculated and observational light curves are shown in Fig.~\ref{lcs2}. 

\subsection{TYC~9356-355-1}
A binary fit with V$_{gfb}$=0.35 in the SED analysis  corresponded to that the temperature of the primary and secondary components are $T_1$=7000~K, $T_2$=4750~K whilst the log~$g$ values were found to be 4.5 for both of the components. The extinction was adopted to 0$\overset{m}{.}$2026 and the distance of the system is 320(4) pc \citep{gai16,gai22}. The orbital period was calculated as 1.31699 by using the method mentioned earlier in the text. 
 
\begin{figure}
\centering
\includegraphics[width=\textwidth]{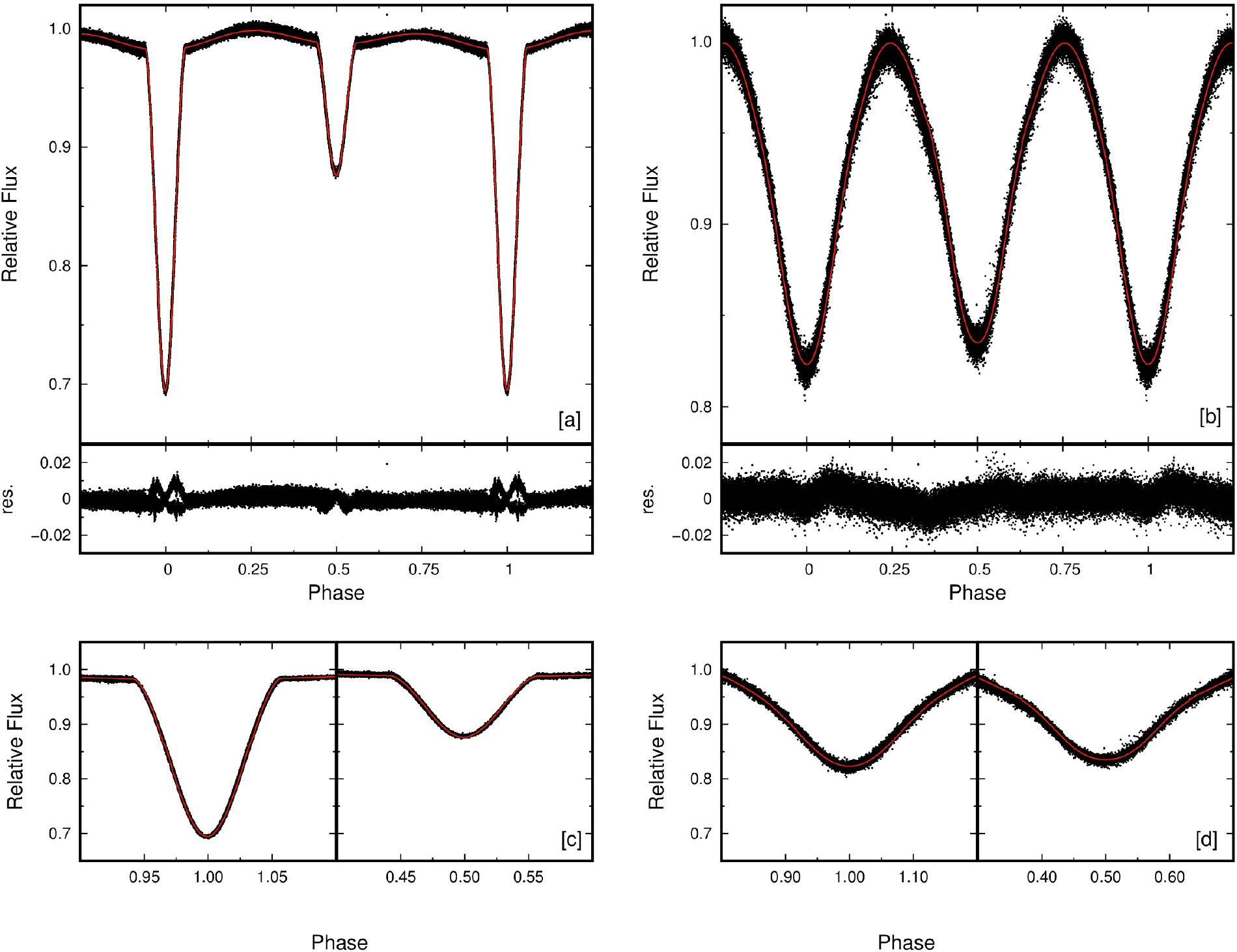}
\caption{Same as \ref{lcs1}, but for TYC~9356-355-1 (a, c) and UCAC4~136-007295 (b, d).}
\label{lcs3}
\end{figure}

The orbital period value, the result of the q-search and the shape of the light curve led us to analyse the light curve, which covers 37859 data points, in detached mode and, therefore, the initial value for the mass ratio was set to 0.5. The calculated light curve was plotted with observations in Fig.~\ref{lcs3} and the light parameters found by the solution are given in Table~\ref{tablc}.

\subsection{UCAC4~136-007295}
The temperatures and the gravities of the components were calculated as  $T_1$=7000~K, $T_2$=4250~K and log~$g_1$=log~$g_2$=4.0 during the SED analysis, the former is relatively close to the value given by \cite{sha18}, 6529~K, in their catalog. The distance of the system is 1295(68)~pc \citep{gai16,gai22} and the extinction through the direction is $A_V$=0$\overset{m}{.}$1195 that yielded using the same method given above. The goodness of fit for the solution, V$_{gfb}$=0.218, corresponds to a good fit.

We analysed the light curve in two morphological assumptions; semidetached (with a secondary filling its Roche lobe) and contact. The semidetached analysis resulted in a warning that the primary component exceeds its corresponding Roche lobe. Thereby, the solution was conducted using 41110 light curve data with an approach that the binary is a contact system. The calculated light curve is plotted with the observations in Fig.~\ref{lcs3} and the results are listed in Table~\ref{tablc}.

%As discussed above, …
%As explained earlier, …
%As previously stated, …
%As indicated previously …
%As described on the previous page, …
%As was mentioned in the previous chapter, …
%thus
%thereby

\section{Conclusion}
We presented a comprehensive study on the binary properties of six eclipsing binary systems. The absolute parameters (Table~\ref{tababs}) are derived based on the light parameters by using the {\tt{AbsParEB}} program \citep{lia15}. It is worth emphasizing that the errors in Table~\ref{tablc} and \ref{tababs} are unphysical since the uncertainties in the Wilson-Devinney method are underestimated. Furthermore, the relatively poor fit caused deviations from the observations around the minimum phases of some systems. The maximum value of the amplitude of deviation was calculated as 0$\overset{m}{.}$02 for CD-58~791 and TYC~8508-1413-1, while it was 0$\overset{m}{.}$03 for CD-62~1257 and TYC~9356-355-1. The masses of the primary components of contact and semidetached systems (CD-54~942, TYC~8508-1413-1 and UCAC4~136-007295) were estimated among 204930 binary star models created by using the Binary Star Evolution code by \cite{hur02,hur13} by setting the initial parameters of the orbital period between 0.5~d and 7~d, the eccentricity between 0 and 0.5, the mass of the primary component, $M_1$, between 0.5 M$_{\odot}$ and 5.0 M$_{\odot}$, the mass of the secondary component between 0.1 M$_{\odot}$ and $M_1$. The masses for the primaries of the detached systems (CD-58~791, CD-62~1257 and TYC~9356-355-1), were calculated from the stellar tracks of \cite{ber09} according to their log $g$ and effective temperature values. The stars are assumed to be in solar abundance during calculations.

\begin{table}
\bc
\begin{minipage}[]{100mm}
\caption[]{Absolute parameters of the systems. TYC~8508, TYC~9356 and UCAC4~136 refer to TYC~8508-1413-1, TYC~9356-355-1 and UCAC4~136-007295, respectively. The standard errors are given in parentheses for the last digits}\label{tababs}\end{minipage}
\setlength{\tabcolsep}{1pt}
\small
 \begin{tabular}{lccc}
  \hline\noalign{\smallskip} 
Parameter & ~~~CD-54~942~~~ & CD-58~791~~~ & CD-62~1257 \\
\hline   
M$_1$ (M$_{\odot}$) 	&0.8		& 1.5		& 2.1		\\
M$_2$ (M$_{\odot}$) 	&0.190(1)	& 1.183(2)	& 1.277(2)	\\
R$_1$ (R$_{\odot}$) 	&1.015(2)	& 2.13(1)	& 2.09(9)	\\
R$_2$ (R$_{\odot}$) 	&0.53(1)	& 1.87(2)	& 1.1(3)	\\
L$_1$ (L$_{\odot}$) 	&0.701(2)	& 7.2(1)	& 6.9(6)	\\
L$_2$ (L$_{\odot}$) 	&0.154(7)	& 5.4(1)	& 1.8(9)	\\
$a $ (R$_{\odot}$) 	& 1.965(2)	& 8.943(4)	& 12.57(1) 	\\
\hline
	& ~~~TYC~8508~~~ & TYC~9356~~~& UCAC4~136 \\
\cline{2-4}                                    
M$_1$ (M$_{\odot}$) 	&1.3		& 1.5		& 1.6		\\
M$_2$ (M$_{\odot}$) 	&0.754(1)	& 0.776(2)	& 0.163(2)	\\
R$_1$ (R$_{\odot}$) 	&1.27(2)	& 1.44(2)	& 1.972(4)	\\
R$_2$ (R$_{\odot}$) 	&1.35(1)	& 1.22(4)	& 0.71(5)	\\
L$_1$ (L$_{\odot}$) 	&2.57(9)	& 4.4(1)	& 8.36(3)	\\
L$_2$ (L$_{\odot}$) 	&1.07(1)	& 1.26(9)	& 0.9(1)	\\
$a $ (R$_{\odot}$) 	&4.074(2)	& 6.807(4)	& 3.382(3) \\
\hline
\end{tabular}
\ec
\end{table}

The fill-out factor, which indicates the degree of contact between components, for the contact binaries CD-54~942 and UCAC4~136-007295 were found 0.19 and 0.42 by using the equation ${f = (\Omega_{i}-\Omega)\mathbin{/}(\Omega_{i} -\Omega_{o})}$ \citep{kal09}, where $\Omega_i$ and $\Omega_o$ are the inner and outer Lagrangian potentials. The filling factor ($f=r/r_{L}$) for the primary component of the semidetached system TYC~8508-1413-1, on the other hand, was calculated by using the relation ${r_{L}=0.49q^{\frac{2}{3}} \mathbin{/} [0.6q^{\frac{2}{3}} + \ln(1+q^{\frac{1}{3}})]}$ given by \cite{egg83}, where $r_L$ is the Roche lobe radius. The $f$ was found to be 0.93 which indicates that the component is close to its Roche lobe and the binary can be considered as a V1010~Oph type near-contact binary group based on the classification of \cite{sha94}. The difference between the maximums of the light curve also supports this classification.

\begin{figure}
\centering
\includegraphics[width=\textwidth]{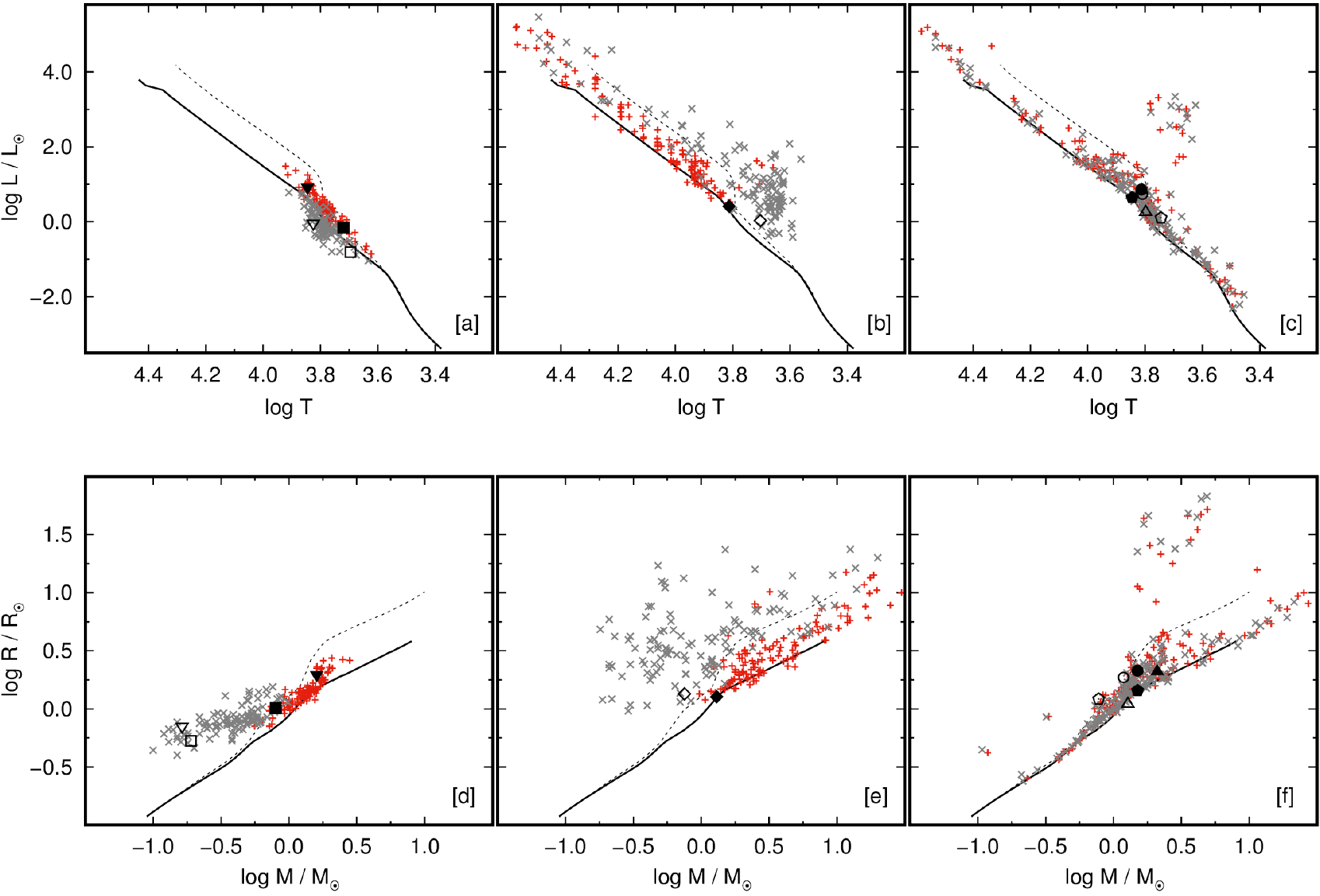}
%\includegraphics[width=0.5\textwidth]{example-image}
%\missingfigure[figwidth=\textwidth]{lc 2617 1499}
\caption{The upper panel shows the Hertzsprung-Russell diagrams while the lower panel indicates the mass-radius planes for contact (a,d), semidetached (b,e) and detached (c,f) binary systems. Plus and cross signs denote the primaries and secondaries of well-known systems with the corresponding morphology. Full and open signs refer to primary and secondary components. Squares, circles, triangles, diamonds, pentagons and upside-down triangles denote CD-54~942, CD-58~791, CD-62~1257, TYC~8508-1413-1, TYC~9356-355-1 and UCAC4~136-007295, respectively. Note that the location of the primary components of CD-58~791 and CD-62~1257 overlap on the Hertzsprung-Russell diagram (c). The data for ZAMS (thick solid line) and TAMS (dashed line) are taken from \citet{bre12} with $Y=0.267$ and $Z=0.01$ abundance.}
\label{hrmr}
\end{figure}

\begin{figure}
\centering
\includegraphics[width=\textwidth]{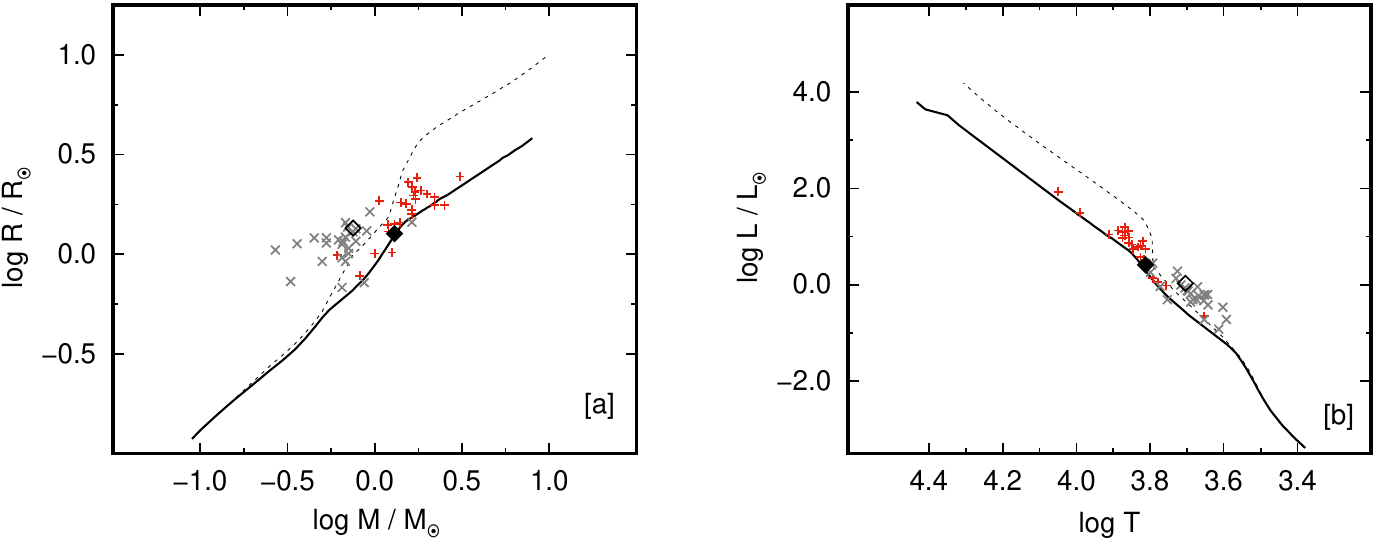}
%\includegraphics[width=0.5\textwidth]{example-image}
%\missingfigure[figwidth=\textwidth]{lc 2617 1499}
\caption{same as Fig.~\ref{hrmr}, but for TYC~8508-1413-1 and the known near-contact binaries. }
\label{ncbhrmr}
\end{figure}

\begin{figure}
\centering
\includegraphics[width=\textwidth]{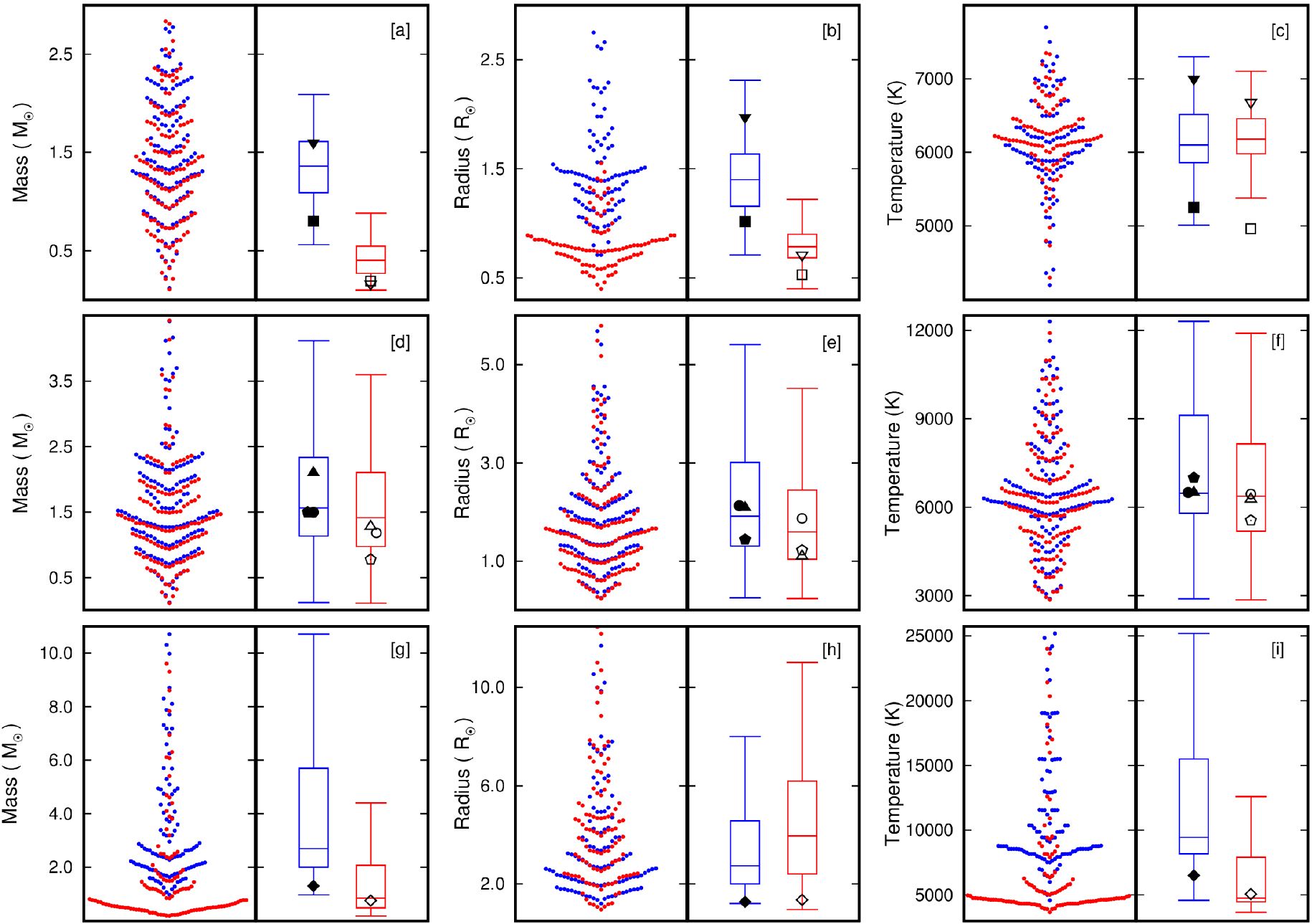}
%\includegraphics[width=0.5\textwidth]{example-image}
%\missingfigure[figwidth=\textwidth]{box all}
\caption{Mass, radius and temperature distributions (left panes) and corresponding box plots (right panes) for primary (blue) and secondary (red) components of known contact (a, b, c), detached (d, e, f) and semidetached (g, h, i) binary systems. The other symbols are the same as those displayed in Fig.~\ref{hrmr}. Some extreme values are excluded in plots, however, boxes are calculated using all available data.}
\label{boxcon}
\end{figure}

The systems were compared to the known binaries with the same morphological type in Fig.~\ref{hrmr}. The plots confirm that the components of the systems are in good agreement with  100 contact \citep{yil13}, 162 detached \citep{sou15} and 119 semidetached \citep{mal20} binary systems. We also compare TYC~8508-1413-1 to 26 known near-contact systems given by \cite{yak05} in Fig.~\ref{ncbhrmr} in order to check its possible near-contact candidacy remarked in the previous paragraph. The components of the target are in accordance with the other near-contact binaries.  A further statistical comparison was made through the known systems' mass, radius and temperature distributions similar to those done by \cite{ula23}. Fig.~\ref{boxcon} illustrates the distributions and the box plots calculated by using the data of corresponding stellar parameters. The  masses, radii and temperatures of our target systems are well inside the minimum and maximum values although the secondary component of  CD-54 942 seems to be cooler than most of its cousins. The parameters of the box plots are given in Table~\ref{tabbox}.

\begin{table}
\bc
\begin{minipage}[]{100mm}
\caption[]{Parameters of box plot calculated by using data from 100 contact \citep{yil13}, 162 detached \citep{sou15} and 119 semidetached \citep{mal20} binaries. $Q1$, $Q2$, $Q3$ and $IQR$ refer to lower quartile, median, upper quartile and interquartile range values.}\label{tabbox}\end{minipage}
\setlength{\tabcolsep}{1pt}
\small
 \begin{tabular}{lcccccc}
  \hline\noalign{\smallskip} 
 & $Q1$~~ &~~ $Q2$~~ &~~ $Q3$~~ &~~ min.~~ & ~~max.~~ &~~ $IQR$\\
 \hline
\underline{Contact} &  & &   &  & & \\
M$_1$ (M$_{\odot}$)&1.09&1.36&1.61&0.56&2.09&0.52\\
M$_2$ (M$_{\odot}$)&0.27&0.40&0.55&0.10&0.88&0.28\\
R$_1$ (R$_{\odot}$)&1.16&1.40&1.58&0.71&2.20&0.43\\
R$_2$ (R$_{\odot}$)&0.69&0.78&0.89&0.40&1.19&0.21\\
T$_1$ (K)	   &5860&6100&6500&5012&7300&640\\ 
T$_2$ (K)	   &5982&6180&6453&5380&7102&472\\  
\hline 
\underline{Detached} &  & &   &  & & \\
M$_1$ (M$_{\odot}$)& 1.13  &   1.56   &  2.33    & 0.12  &   4.12    & 1.19 \\
M$_2$ (M$_{\odot}$)& 0.97   &  1.41    & 2.10    & 0.11  &   3.60    & 1.13 \\
R$_1$ (R$_{\odot}$)&1.31    & 1.92    & 2.98    & 0.25  &   5.41    & 1.67  \\   
R$_2$ (R$_{\odot}$)&1.04     &1.59    & 2.43    & 0.24  &   4.29    & 1.38  \\   
T$_1$ (K)&5798 & 6471 & 9094 & 2892 & 12303 &3296 \\
T$_2$ (K)&5191  &6375 & 8133 & 2851 & 11912& 2942 \\
\hline 
\underline{Semidetached} &  & &   &  & & \\
M$_1$ (M$_{\odot}$) 	&2.01	& 2.71	& 5.69 &0.97 &10.70 & 3.68	\\
M$_2$ (M$_{\odot}$) 	&0.48   & 0.88  & 2.03 &0.18 &4.10  &1.55	\\
R$_1$ (R$_{\odot}$) 	&2.00	& 2.74  & 4.50 &1.20 &8.00  &2.50	\\
R$_2$ (R$_{\odot}$) 	&2.41	& 4.01	& 6.12 &0.96 &11.02 &3.70	\\
T$_1$ (K) 	            &8283	& 9660	& 15498 & 4570 &25200 &7215	\\
T$_2$ (K) 	            &4465	& 4803	& 7615	&3650 &10900 &3150 \\
\hline
\end{tabular}
\ec
\end{table}

We concluded that TYC~8508-1413-1 is a semidetached system with the probability of being a V1010~Oph type near-contact system. CD-54~942 and UCAC4~136-007295 are contact binaries and CD-58~791, CD-62~1257 and TYC~9356-355-1 are detached systems. Spectroscopic studies in combination with our results may lead to determining the characteristics of the systems more precisely.

\normalem
\begin{acknowledgements}
This work was supported by \c{C}anakkale Onsekiz Mart University, The Scientific Research Coordination Unit, Project number: 2022-FBA-3851. The numerical calculations reported in this paper were partially performed at TUBITAK ULAKBIM, High Performance and Grid Computing Center (TRUBA resources). This paper includes data collected by the TESS mission. Funding for the TESS mission is provided by the NASA's Science Mission Directorate. Some/all of the data presented in this paper were obtained from the Mikulski Archive for Space Telescopes (MAST). STScI is operated by the Association of Universities for Research in Astronomy, Inc., under NASA contract NAS5-26555. Support for MAST for non-HST data is provided by the NASA Office of Space Science via grant NNX13AC07G and by other grants and contracts. This research has made use of the NASA/IPAC Infrared Science Archive, which is operated by the Jet Propulsion Laboratory, California Institute of Technology, under contract with the National Aeronautics and Space Administration. This publication makes use of VOSA, developed under the Spanish Virtual Observatory project supported by the Spanish MINECO through grant AyA2017-84089. VOSA has been partially updated by using funding from the European Union's Horizon 2020 Research and Innovation Programme, under Grant Agreement nº 776403 (EXOPLANETS-A). This research has made use of NASA's Astrophysics Data System. This research has made use of the VizieR catalogue access tool, CDS, Strasbourg, France.This work has made use of data from the European Space Agency (ESA) mission
{\it Gaia} (\url{https://www.cosmos.esa.int/gaia}), processed by the {\it Gaia}
Data Processing and Analysis Consortium (DPAC,
\url{https://www.cosmos.esa.int/web/gaia/dpac/consortium}). Funding for the DPAC
has been provided by national institutions, in particular the institutions
participating in the {\it Gaia} Multilateral Agreement.

\end{acknowledgements}
  
\bibliographystyle{raa}
\bibliography{Ulas_bib}

\end{document}